\documentstyle[12pt,axodraw]{article}
\newcommand{\Fh}[2]{\,{}_#1F_#2}
\newcommand{\Fs}[3]{\!\!\left[\begin{array}{c}#1\,;\\#2\,;\end{array}#3\right]}

\newcommand{\Ffp}[2]{\Fs{#1}{#2}{\frac{q^2}{4m^2}}}

\newcommand{\Ffr}[2]{\Fs{#1}{#2}{1-\frac{m_1^2}{m_2^2}}}

\begin{document}
\thispagestyle{empty}
\onecolumn
\date{}
\vspace{-1.4cm}
\begin{flushleft}
{DESY 96-099 } \\
{JINR E2-96-174 }\\
{hep-ph/9606238} \\
June,~1996

\end{flushleft}
\vspace{1.5cm}

\begin{center}
{\LARGE {\bf

A new approach to the momentum

expansion of  multiloop Feynman

diagrams}

}
\vfill

{\large
 O.V.Tarasov \footnote{E-mail:
 $tarasov@ifh.de$ \\
 {~On leave of absence from JINR, 141980 Dubna (Moscow Region),
 Russian Federation.}}}

\vspace{2cm}
Deutsches Electronen-Synchrotron DESY \\
Institut f\"ur Hochenergiephysik IfH, Zeuthen\\
Platanenallee 6, D--15738 Zeuthen, Germany
\vspace{1cm}
\end{center}

\vfill

\begin{abstract}
We present a new  method for the momentum expansion of Feynman 
integrals with arbitrary masses and any number of loops and external
momenta. By using the parametric representation we derive a 
generating function for the coefficients of the small momentum 
expansion of an arbitrary diagram. The method  is  applicable for the
expansion w.r.t. all or a subset of external momenta. The 
coefficients of the expansion are obtained by applying a differential
operator to a given integral with shifted value of the  space-time 
dimension $d$ and the expansion momenta set equal to zero. Integrals
with changed  $d$ are evaluated by using the generalized recurrence
relations  proposed in  \cite{OVT1}. We show how the method works for
one- and two-loop integrals. It is also illustrated  that our method
is simpler and more efficient than others.
\end{abstract}

\vfill
\newpage

\setcounter{footnote}{0}
\section{Introduction}

Accurate theoretical predictions for precision experiments in particle
physics are usually  the result  of  formidable
calculations of radiative corrections.  More and more  particles with
different masses have to be taken into account. This requires
the evaluation of complicated  Feynman integrals with many masses and 
external
momenta. Under lucky circumstances one can get analytic results for
the radiative corrections but more frequently one is forced to
expand w.r.t. external momenta and/or masses or to find other kinds
of approximate calculations  \cite{3loopQED}, \cite{Czarnecki}.
Frequently even if the result for a particular diagram is known
analytically \cite{Buza}, the numerical evaluation proceeds via
approximations which again may be some kind of momentum
expansion. It is evident that methods of asymptotic expansions
of Feynman integrals \cite{asymptotic} will play an increasing role
in the evaluation of physical amplitudes.

The asymptotic momentum expansions combined with conformal mapping
and subsequent resummation by means of Pad\`e approximations
were used to calculate the diagrams in the whole cut plane in
the external momentum squared \cite{FT}-\cite{FST}. As was found in 
these papers  the method gives very precise
results in a wide range of  kinematical variables.
  To extend  further  this method  we need
to find a systematic and effective method for the momentum expansion
of diagrams depending on several external momenta and masses.
For the three-point Green functions this problem was solved in 
\cite{OVT2}, \cite{DT1}. As for diagrams with more external legs no
reasonable algorithm is known up to now.

The solution of this problem is important not only for the case of
small momenta. Asymptotic expansions at large momenta or masses
also include  small momentum expansion as an ingredient part
\cite{DTS}.

In the present paper we describe a new approach to the momentum
expansion which solves the aforementioned problem.
The method is closely related to the new, generalized
recurrence relations for Feynman integrals proposed in \cite{OVT1}.

The paper is organized as follows.
In Sect.2, we present the main ideas of our method.
First, using the parametric representation, we give a derivation
of the generating function  for the  coefficients of the expansion
of an arbitrary propagator type  integral with different masses.
Then we derive a similar  formula for the diagrams with arbitrary
number of external legs.

In Sect.3 we demonstrate how the method works for one- loop
integrals. A general formula,  in terms of a one-fold integral,
for the coefficients of the expansion of scalar n-point one-loop
diagrams is derived.

In Sect.4 several examples of the small momentum  expansion of 
various two-loop diagrams are considered. We show how to expand 
integrals w.r.t. subsets of external momenta. The characteristic 
functions required for  expanding propagator type  and 
three-point vertex  diagrams are given.


\section{ General formalism }


 The subject of our consideration will be dimensionally
regulated scalar Feynman integrals.
An arbitrary scalar $L$ loop integral can be written as
\begin{equation}
G^{(d)}(\{s_i\},\{m_s^2\})=\prod_{i=1}^{L} \int d^dk_i \prod^{N}_{j=1}
P^{\nu_j}_{\overline{k}_j,m_j},
\label{arbdia}
\end{equation}
where
\begin{equation}
P^{\nu}_{k,m}=\frac{1}{(k^2-m^2+i \epsilon)^{\nu}},
~~~~~~~~~~~~~~~~~\overline{k}_j^{\mu}=
 \sum^L_{n=1} \omega_{jn}k^{\mu}_n+ \sum _{m=1}^E
 \eta_{jm}q_m^{\mu},
\end{equation}
$q_m$ are external momenta, $\{s_i\}$
is a set of scalar invariants formed from $q_m$, $N$ is the number of lines,
$E$ is the number of external legs, $\omega$ and $\eta$ are matrices
of incidences of the graph with the matrix elements being $\pm1$ or $0$
(see, for example, Ref. \cite{IZ}).

In what follows we will need a parametric representation
for $G^{(d)}(\{s_i\},\!\{m_s^2\})$.
For an arbitrary scalar Feynman
integral in $d$ dimensional space-time we have \cite{IZ},
\cite{parametric}:
\begin{equation}
G^{(d)}(\{s_i\},\!\{m_s^2\})=i^L\! \left ( \frac{\pi}{i} \right)^{
\frac{dL}{2}} \prod^{N}_{j=1} \frac{i^{-\nu_j}}{\Gamma(\nu_j)}
 \int_0^{\infty} \!\!\!\! \ldots \! \int_0^{\infty} \!\frac{  d 
 \alpha_j \alpha^{\nu_j-1}_j}{ [ D(\alpha) ]^{\frac{d}{2}}}
 e^{i[\frac{Q(\{s_i\},\alpha)}{D(\alpha)}
 -\sum_{l=1}^{N}\alpha_l(m_l^2-i\epsilon)]},
\label{DQform}
\end{equation}
where   $D(\alpha)$
and $Q(\{s_i\},\alpha)$ are homogeneous polynomials
in $\alpha$ of degree $L$ and $L+1$, respectively.
They can be represented as sums over trees and
two-trees of the graph  (see, for example, Ref.\cite{BS}):
\begin{eqnarray}
D(\alpha)&\!=&\!\!\!\sum_{
\begin{array}{l}
{\rm over}\\{\rm trees}
\end{array}} ( \prod_{\begin{array}{l}{\rm over}\\{\rm chords}
\end{array}}\!\!  \ldots \alpha_j \ldots \!\!), \nonumber \\
&&\nonumber \\
Q(\{s_i\},\alpha)&\!=&\!\!\!\sum_{\begin{array}{l}
 {\rm ~over} \\{\rm 2\!-\!trees}
 \end{array}} ~~ \prod_{\begin{array}{l}
 {\rm ~over}\\{\rm chords}
 \end{array}} \!\!
\ldots \alpha_j \ldots \!\!~~ (\sum_{
\begin{array}{l}{ \rm over~comp.}\\ {\rm of~2\!-\!tree}
\end{array}} q~)^2.
\end{eqnarray}
These polynomials are characteristic functions of the topology
of the diagram and of its subgraphs.
Since $D$ and $Q$ will play an important role in the rest of the
present paper we remind the reader of the definitions of the trees
and two-trees for   connected diagrams.
Any connected subdiagram of a diagram $G$ containing all the
vertices of $G$ but is free of cycles (loops) is called a tree of
$G$. Similarly, a two-tree is defined as any subdiagram of $G$
containing all the vertices of the original diagram,
but is free of cycles, and consisting of exactly two connected
components. Finally, a chord of a tree (two-tree) is
defined as any line not belonging to this tree (two-tree).

For the practical application of our method the following properties
of the functions $D$ and $Q$ are important. Namely, the dependence
of $D$ and $Q$ on each $\alpha_{\nu}$, can be written
in the form:
\begin{eqnarray}
&&D(\alpha)=D_{\overline{\nu}}(\alpha) \alpha_{\nu}+
D_{\nu}(\alpha), \nonumber \\
&&Q(\{s\},\alpha)=Q_{\overline{\nu}}(\{s\},\alpha) \alpha_{\nu}+
Q_{\nu}(\{s\},\alpha),
\label{DQshrink}
\end{eqnarray}
where
$D_{\overline{\nu}}(\alpha), ~Q_{\overline{\nu}}(\{s\},\alpha)$
are the $D$, $Q$ functions of the diagram obtained by removing the
line $\nu$, and $D_{\nu}(\alpha),~Q_{\nu}(\{s\},\alpha)$
are the $D, ~Q$ functions for the diagram obtained by contracting
the line $\nu$ to a point. $D_{\nu}$, $D_{\overline{\nu}}$, $Q_{\nu}$ 
and $Q_{\overline{\nu}}$ do not depend on $\alpha_{\nu}$.
The polynomials $D(\alpha)$, $Q(\{s_i\},\alpha)$ can be easily
constructed by means of a computer program  for any particular 
integral \cite{OVT}. Examples of $D$ and $Q$ for some particular 
diagrams will be given in the next sections.

Let us first  consider  propagator type integrals. In this case $Q$
is proportional to the external momentum squared $q^2$: 
$Q(q^2,\alpha)=\overline{Q}(\alpha)q^2$.
Assuming temporarily that all propagators have different masses,
from the  representation ($\ref{DQform}$) we obtain:
\begin{equation}
\frac{\partial}{\partial q^2} G^{(d)}(q^2,\{ m^2_s\})=
\frac{(-1)^{L+1}}{\pi^L}  \overline{Q}
\left( \partial \right) G^{(d+2)}
( q^2,\{ m^2_s \}),
\end{equation}
where $\overline{Q}\left( \partial \right)$, is the differential 
operator obtained from $\overline{Q}(\alpha)$ by replacing
$\alpha_j \rightarrow \partial_j \equiv \partial / \partial m_j^2$.
In the above formula it is assumed that after 
differentiation all  masses are set  to their original value.

Thus,   differentiation w.r.t. momenta of the $d$ dimensional
integral is replaced by  differentiation  w.r.t. masses
of the $d+2$ dimensional one. As we shall show in the next sections
this replacement in many cases will essentially simplify practical
calculations. For the $l$-th derivative we can write
\begin{equation}
\left( \frac{\partial}{\partial q^2}\right)^l G^{(d)}(q^2,\{ m^2_s\})
=\frac{(-1)^{(L+1)l}}{\pi^{Ll}}
\overline{Q}^l (\partial) G^{(d+2l)}(q^2,\{ m^2_s\}).
\label{equation7}
\end{equation}
Upon setting $q^2=0$ we obtain the coefficients of the small momentum
expansion of the propagator type diagram.

The general prescription for the small momentum expansion
is then  the following.
First, one should assign to all lines different masses
and then set the external momenta equal to zero. Second, 
the differential operator $Q(\partial)$ must be applied  as much as
needed and then the masses must be set to the  original one's.
Third, the integration by parts recurrence relations in $d+2l$
dimensions should be used, reducing all the integrals to
the set of master integrals.
Fourth, the set of all master integrals in $d+2l$ dimensions must
be expressed in terms of $d$ dimensional master integrals
using the generalized recurrence relations described in \cite{OVT1}.
In some cases it is advantageous  to use recurrence relations
which simultaneously  reduce the value of $d$ and the
sum of exponents of the propagators in the integral.
Rather frequently  master integrals are just combinations of
Euler's $\Gamma$ functions, and   transforming the latter
from $d+2l$ to $d$ dimensions is quite simple.

In practical calculations no real differentiation is needed.
It is enough to compute $Q^l(\alpha)$, multiply it by
$G^{(d+2l)}(0,\{ m^2_s\})$
and make the substitutions
$$
\frac{\alpha^j_i}{[k^2_s-m_i^2+i\epsilon]^n}=\frac{\Gamma(n+j)}
{\Gamma(n)[k^2_s-m_i^2+i\epsilon]^{n+j}}.
$$

 An important feature of our method is that one does not need to
care about external momenta, avoiding many complications intrinsic
to  other methods based on the differentiation w.r.t. external
momenta. From the very beginning, one is dealing with simpler
objects; only  bubble diagrams need to be  considered.
In all  cases where the bubble diagrams  are known,
one immediately gets an explicit generating function for the
coefficients.

Unfortunately the polynomial $Q$ usually has many terms,  and it is
not so easy to exponentiate it. For the three-loop nonplanar
master  integral occurring in the QED photon propagator,
$Q$ is a fourth order polynomial with 45 terms. In this case
using FORM \cite{FORM} running on a DEC 3000 it took about
46 hours  to generate the  12-th coefficient in the small momentum
expansion in terms of bubble integrals.
Nevertheless, the number of generated terms, and as a consequence
the execution time, is smaller than what one gets using
the direct small momentum expansion methods. Using the method
proposed recently in \cite{Baikov} for solving
recurrence relations for three-loop vacuum integrals one may expect
an improvement of the efficiency in the complete calculations of the
coefficients.
Our  method is definitely  the most optimal for the case
of  all different masses.
In case when some masses in the integral are equal the
calculation can be  simplified by taking  into account this
information from the very beginning.

The algorithm described above can be generalized to the small
momentum expansion of  integrals with an arbitrary number of external
momenta. For integrals depending on several external momenta the
polynomial $Q$ is a sum over a set of $n=E(E-1)/2$ independent scalar
invariants $s_j$:
\begin{equation}
Q(\alpha)= \sum_{j=1}^n s_j Q_j( \alpha),
\end{equation}
where $E$- is the number of external legs of the diagram
corresponding  to a given integral. In this case the analogue
of ($\ref{equation7}$), valid for propagators, reads:
\begin{equation}
 \frac{\partial^l}{\partial s^l_j} G^{(d)}(\{s_k \},\{ m^2_i\})
=\frac{(-1)^{(L+1)l}}{\pi^{Ll}}
Q^l_j (\partial)G^{(d+2l)}(\{s_k \},\{ m^2_i\}).
\label{siequation}
\end{equation}

To find the coefficient of the $M$-th order in
$ s^{l_1}_1\ldots s^{l_n}_n$ one should apply the product of
operators  $ Q^{l_1}_1(\partial) \ldots Q^{l_n}_n(\partial) $
to the integral taken in $d+2M$ dimensions. If we perform an
expansion  w.r.t.  all scalar  invariants again  the coefficients
of the expansion will be generated from the bubble
integrals. In this case the general formula for the small momentum
expansion of an arbitrary scalar integral will be:
\begin{eqnarray}
&&G^{(d)}(\{s_k\},\{m_i^2 \})=
\sum_{M=0}^{\infty} \frac{(-1)^{M(L+1)}}{\pi^{ML}}
\nonumber\\
&&~~~~~~~~\times \sum_{\{l_1+\ldots  +l_n=M\}}
 Q^{l_1}_1(\partial)
\ldots  Q_n^{l_n}(\partial)
G^{(d+2M)}(\{ 0 \},\{m_i^2 \})\frac{s_1^{l_1}}{l_1!} \ldots
\frac{s_n^{l_n}}{l_n!}.
\label{EXPANSION}
\end{eqnarray}

The expansion can also be performed  w.r.t. some subset of scalar
invariants which are proportional to the momenta considered to be small.
In this case the operators $Q_j(\partial)$
must be applied to the given  integral where only that
subset of the momenta is taken to be zero. The corresponding
formula is almost identical to ($\ref{EXPANSION}$) and  may be
readily written as:

\begin{eqnarray}
&&G^{(d)}(\{s_k\},\{S_F\},\{m_i^2 \})=
\sum_{M=0}^{\infty} \frac{(-1)^{M(L+1)}}{\pi^{ML}}
\nonumber\\
&&~~~~~~~~\times \sum_{\{l_1+\ldots  +l_n=M\}}
 Q^{l_1}_1(\partial)
\ldots  Q_n^{l_n}(\partial)
G^{(d+2M)}(\{ 0 \},\{S_F\},\{m_i^2 \})\frac{s_1^{l_1}}{l_1!} \ldots
\frac{s_n^{l_n}}{l_n!},
\label{EXPANSION2}
\end{eqnarray}
where $\{S_F\}$ is a set of scalar invariants assumed to be fixed.

Equations  ($\ref{EXPANSION}$) and ($\ref{EXPANSION2}$) can be used
not only  for the small momentum expansion. The momentum
expansion  considered is also an essential ingredient  of
other kind of asymptotic expansions \cite{FST}, \cite{DTS},  \cite{BDS}.

\section{Expansion of one-loop integrals}
Now we shall show how the method works by considering  several simple
examples. We start with the one-loop integral:
\begin{equation}
I^{(d)}_{\alpha \beta}(q^2,m_1^2,m_2^2)=\int \frac{d^d k_1}
{[i \pi^{d/2}]} P^{\alpha}_{k_1,m_1} P^{\beta}_{k_1-q,m_2}
= \sum_{l=0}^{\infty} \frac{I_l}{l!}(q^2)^l.
\end{equation}
In this case $Q(q^2,\alpha)=\alpha_1 \alpha_2 ~q^2$.
The $l$-th coefficient of the small momentum expansion will be
\begin{equation}
I_l=\frac{\partial^l}{\partial (m_1^2)^l}
\frac{\partial^l}{\partial (m_2^2)^l}
 \int \frac{d^{d+2l}k_1}{[i\pi^{d/2+l}]} ~ P_{k_1,m_1}^{\alpha}
 P_{k_1,m_2}^{\beta}.
\end{equation}
Upon differentiation we obtain:
\begin{equation}
I_l= \frac{\Gamma(l+\alpha)~\Gamma(l+\beta)}
         {\Gamma(  \alpha)~\Gamma(  \beta) }
\int \frac{d^{d+2l} k_1~}{[i \pi^{d/2+l}]} P_{k_1,m_1}^{\alpha+l}
P_{k_1,m_2}^{\beta+l}.
\end{equation}
The bubble integral can be expressed in terms of hypergeometric
function $~_2F_1$:
\begin{equation}
\int \frac{d^dk_1}{[\pi^{d/2}]} P^{\alpha}_{k_1,m_1}
 P^{\beta}_{k_1,m_2}=
\frac{(-1)^{\alpha+\beta}\Gamma \left(\alpha+\beta-\frac{d}{2}\right)}
{(m_1^2)^{\alpha-\frac{d}{2}} (m_2^2)^{\beta}~\Gamma(\alpha+\beta)}
 \Fh21\Ffr{\beta,\frac{d}{2}}{\alpha+\beta}.
\end{equation}
For simplicity we will take $m_1^2=m_2^2=m^2$.
In this case
\begin{equation}
I_l=(-1)^{\alpha+\beta}\frac{\Gamma(l+\alpha)~\Gamma(l+\beta)}
         {\Gamma(  \alpha)~\Gamma(  \beta)}
 	   \frac{\Gamma(\alpha+\beta+l-\frac{d}{2})
	   }{\Gamma(\alpha+\beta+2l) (m^2)^{l+\alpha+\beta-d/2}}.
\end{equation}
and therefore
\begin{equation}
I^{(d)}_{ \alpha \beta}(q^2,m^2,m^2)=
\frac{(-1)^{\alpha+\beta} \Gamma(\alpha+\beta-\frac{d}{2})}
{(m^2)^{\alpha+\beta-\frac{d}{2}}
 \Gamma(\alpha+\beta)}
 \Fh32\Ffp{\alpha,\beta,\alpha+\beta-\frac{d}{2}}
  {\frac{\alpha+\beta}{2},\frac{\alpha+\beta+1}{2}},
\label{f32}
\end{equation}
in agreement  with Ref.\cite{BoDa}.

To compare our method with  \cite{DT} it is instructive to calculate
the coefficient $I_1$ at $m_1^2=m_2^2=m^2$,
$\alpha=\beta=1$ using both techniques.
In our approach $I_1$ is just one term:
\begin{eqnarray}
I_1&=& \frac{\partial^2}{\partial m_1^2~\partial m_{2}^2}
\int \left. \frac{d^{d+2}k_1}{[i\pi^{d/2+1}]} \frac{1}{(k_1^2-m_1^2+i 
\epsilon)(k_1^2-m_2^2+i\epsilon)} \right|_{m_1^2=m_2^2=m^2}  
\nonumber \\
& & \nonumber \\
&=& \int \frac{d^{d+2}k_1}{[i\pi^{d/2+1}]} \frac{1}{(k_1^2-m^2
+i\epsilon)^4}.
\end{eqnarray}
The same coefficient in the method \cite{DT} is given by
\begin{eqnarray}
I_1&=&
\frac{1}{2d} \frac{\partial^2}{\partial q_{\mu} \partial q^{\mu}}
\int \left. \frac{d^{d}k_1}{[i\pi^{d/2}]} \frac{1}{((k_1-q)^2-m^2
+i\epsilon)
(k_1^2-m^2+i\epsilon)} \right|_{q=0} \nonumber \\
& & \nonumber \\
&=&\frac{1}{d} \int \frac{d^{d}k_1}{[i\pi^{d/2}]} ~\frac{[4k_1^2
-d(k_1^2-m^2)]}{(k_1^2-m^2+i\epsilon)^4}.
\label{lapla}
\end{eqnarray}
One can observe at first, that  differentiating w.r.t. masses is 
easier than w.r.t. external momenta.  Secondly, after differentiation
the  number of terms for $I_1$ in ($\ref{lapla}$) is lager  than in 
our approach. The number of terms for the higher coefficients $I_l$ 
in the approach \cite{DT} grows exponentially. A very similar 
situation we find for integrals with more loops.

For $d$ dimensional one-loop integrals corresponding  to
diagrams with $E$ legs the number of terms in $Q$
is equal to the number of independent scalar invariants $n=E(E-1)/2$.
In this case $Q$ may be represented as:
\begin{equation}
Q(\{s\},\alpha)=\sum_{j=1}^{n} Q_j(\alpha) s_j,
\end{equation}
with $Q_j(\alpha)$ being monomials in $\alpha$.
In the small momentum expansion of the one-loop
integrals:
\begin{eqnarray}
G^{(d)}(\{s_i\},\{m_s^2\})&=&
\int d^d k_1 P_{k_1-q_1,m_1}^{\nu_1}\ldots P_{k_1-q_{E-1},m_{E-1}}
^{\nu_{E-1}} P_{k_1,m_E}^{\nu_E}\nonumber \\
&=&
\sum_{M=0}^{\infty} \sum_{\{ l_1+\ldots +l_n=M \} }
C_{l_1\ldots l_n} \frac{s_1^{l_1}}{l_1!} \ldots
\frac{s_n^{l_n}}{l_n!},
\end{eqnarray}
the coefficients of the expansion will be
\begin{equation}
C_{l_1 \ldots l_n}=\frac{1}{\pi^M}
Q^{l_1}_1(\partial)\ldots Q^{l_n}_n(\partial)
\int d^{d+2M}k_1~
P_{k_1,m_1}^{\nu_1}\ldots P_{k_1,m_E}^{\nu_E}.
\label{equation19}
\end{equation}
Performing the angular integration we obtain a compact one-fold
integral representation for the coefficients of the momentum
expansion of an arbitrary one-loop integral:
\begin{equation}
C_{l_1 \ldots l_n}= \frac{i \pi^{\frac{d}{2}-M}
Q^{l_1}_1(\partial)\ldots Q^{l_n}_n(\partial)
 }{\Gamma(\frac{d}{2}+M)}
\int_0^{\infty} \frac{(-1)^{\nu_1+\ldots +\nu_E} ~x^{\frac{d}{2}
+M-1}~dx}{(x\!+\!m_1^2)^{\nu_1}\ldots (x\!+\!m_E^2)^{\nu_E}}.
\label{equation20}
\end{equation}
The previous integral may readily be evaluated  in terms of the 
Lauricella function $F^{(E-1)}_D$ (see formula 7.1.1.5 in Ref. 
\cite{Exton}) in agreement with Ref. \cite{And1loop}. In the case when all 
masses are equal or some masses are zero while the others are equal 
the integral can be evaluated in terms of Euler's $\Gamma$ function.

For illustration, we consider  the expansion of the three-point
vertex function w.r.t. external momenta $q_1^2$, $q_2^2$, 
$q_3^2=(q_1+q_2)^2$:
\begin{equation}
I_{\nu_1\nu_2\nu_3}=\int \frac{d^dk_1}{[i \pi^{d/2}]}~
P_{k_1-q_1,m_1}^{\nu_1}
P_{k_1,m_2}^{\nu_2}
P_{k_1+q_2,m_3}^{\nu_3}=\sum_{l_1,l_2,l_3=0}^{\infty}
C_{l_1 l_2 l_3} \frac{(q_1^2)^{l_1}(q_2^2)^{l_2}(q_3^2)^{l_3}}
{l_1! ~l_2! ~l_3!}.
\end{equation}
The function  $Q$ corresponding to $I_{\nu_1\nu_2\nu_3}$ reads:
\begin{equation}
Q(\{s\},\alpha)=\alpha_1 \alpha_2 q_1^2+\alpha_2 \alpha_3 q_2^2
+\alpha_1 \alpha_3 q_3^2.
\end{equation}
The coefficients of the expansion in $q_1^2,q_2^2,q_3^2$ may
readily be found from ($\ref{equation19}$) and ($\ref{equation20}$):
\begin{eqnarray}
C_{l_1 l_2 l_3}&\!\!=\!\!&\!
\left(\partial_1 \partial_2 \right)^{l_1}
                \left(\partial_2 \partial_3 \right)^{l_2}
                \left(\partial_1 \partial_3 \right)^{l_3}
\int \frac{d^{d+2l_1+2l_2+2l_3}k_1}{[i \pi^{d/2+l_1+l_2+l_3}]}~
P_{k_1,m_1}^{\nu_1}P_{k_1,m_2}^{\nu_2} P_{k_1,m_3}^{\nu_3}
\nonumber \\
&& \nonumber \\
&\!\!=\!\!&\!\!
\frac{(\nu_1)_{l_1\!+l_3}(\nu_2)_{l_1\!+l_2}(\nu_3)_{l_2\!+l_3}}
{\Gamma(\frac{d}{2}+l_1+l_2+l_3)}
\!\! \int_0^{\infty} \!\!\!\! \frac{(-1)^{\nu_1+\nu_2+\nu_3}~
x^{\frac{d}{2}+l_1+l_2+l_3-1} dx}
{(\!x\!+\!m_1^2)^{\nu_1\!+l_1\!+l_3}(\!x\!+\!m_2^2)^{\nu_2\!
+l_1\!+l_2}(\!x\!+\!m_3^2)^{\nu_3\!+l_2\!+l_3}},
\end{eqnarray}
where $(\nu)_l\equiv \Gamma(\nu+l)/\Gamma(\nu)$ is the Pochhammer 
symbol. For the  case of equal masses $m_1^2=m_2^2=m_3^2=m^2$, the 
coefficients are just products of Euler's $\Gamma$ functions:

\begin{equation}
C_{l_1 l_2 l_3}=
(-1)^{\nu_1+\nu_2+\nu_3} \frac{(\nu_1)_{l_1+l_3}(\nu_2)_{l_1+l_2}
(\nu_3)_{l_2+l_3}
\Gamma \left(\nu_1+\nu_2+\nu_3+l_1+l_2+l_3-\frac{d}{2}\right)}
{(m^2)^{\nu_1+\nu_2+\nu_3+l_1+l_2+l_3-\frac{d}{2}}
~\Gamma(\nu_1+\nu_2+\nu_3+2(l_1+l_2+l_3))}.
\end{equation}

\vspace{0.5cm}

Let us consider now an integral  with  one small momentum, and the
other ones  arbitrary. This kind of kinematics occurs, for instance, in 
the evaluation of the moments of the deep inelastic structure 
functions \cite{BBDM}. The one-loop  scalar diagram which is of 
interest in this case is shown in Fig.1.

\begin{picture}(200,155)(0,5)
\Text(160,140)[]{1}
\Text(210,90)[]{2}
\Text(160,60)[]{3}
\Text(110,90)[]{4}
\Text(105,60)[]{$p$}
\Text(215,60)[]{$p$}
\Text(105,140)[]{$q$}
\Text(215,140)[]{$q$}
\Line(120,50)(200,50)
\Line(120,130)(200,130)
\Line(120,50)(120,130)
\Line(200,50)(200,130)
\ArrowLine(200,50)(230,50)
\ArrowLine(200,130)(230,130)
\ArrowLine(90,50)(120,50)
\ArrowLine(90,130)(120,130)
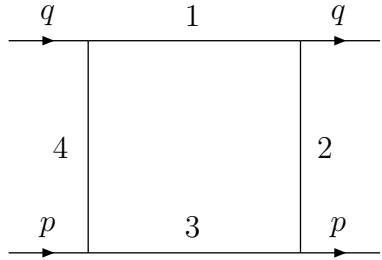
\Text(169,30)[]{${\rm Fig.1 ~Box ~diagram~depending ~on ~two~momenta}$}
\end{picture}

 The contribution to the moments of the structure functions
 from the  integral related to the diagram in Fig.1 will be
 determined by  the coefficients of the expansion  w.r.t. the
 scalar product $(pq)$ with the assumption  $p^2=0$:
\begin{equation}
T(p,q)=\int \frac{d^dk_1}{[i \pi^{d/2}]}
P^2_{k_1,m}P_{k_1+q,m}P_{k_1-p,m}=\sum_{n=0}^{\infty}(pq)^n  T_n .
\end{equation}
From the function $Q$ of the diagram we derive the differential operator 
that generates the coefficients $T_n$:
\begin{equation}
Q_{pq}(\partial)=2\partial_1 \partial_3.
\end{equation}
Introducing auxiliary masses,
setting $p=0$, applying $Q_{pq}^l (\partial)$ to the diagram and
setting  $m_1^2=m^2_2=m^2_3=m^2_4=m^2$  after differentiation,
we get:
\begin{eqnarray}
T_n\!&=&\!2^n n!  \int
\frac{d^{d+2n}k_1} {[i \pi^{d/2+n}]}
P^{n+3}_{k_1,m}P^{n+1}_{k_1+q,m} \nonumber \\
\!&=&\!\frac{2^n n! ~\Gamma \left(n+4-\frac{d}{2}\right)}
{(m^2)^{n+4-d/2}}
\Fh32\Ffp{n+3,n+1,n+4-\frac{d}{2}}
          {n+2,n+\frac{5}{2}}.
\label{equation30}
\end{eqnarray}
The integral in ($\ref{equation30}$) was evaluated  with the aid  of
($\ref{f32}$). A different method \cite{GLT} to calculate moments of
this kind of diagram, consists in applying a differential projection
operator built
up  from the momenta $p$ and $q$. From the above comparison made for
the integral  $I_{\alpha \beta}^{(d)}(q^2,m_1^2,m_2^2)$, the analysis 
given  in the next section for the two-loop diagrams and the remark
in Sect.2  concerning three-loop calculations,  we may expect 
our method to be more efficient than the method of projections.

\section{Expansion of two-loop integrals}

Small momentum expansions of various kinds of two-loop  diagrams
have been considered in \cite{FT}--\cite{FST},
\cite{DT}. For the generic two-loop propagator type diagram
\begin{equation}
J_{\nu_1 \nu_2 \nu_3 \nu_4 \nu_5}(q^2)=
\int \!\! \int \frac{d^dk_1 ~d^dk_2}{[i\pi^{d/2}]^2}
~P^{\nu_1}_{k_1,m_1}P^{\nu_2}_{k_2,m_2}
P^{\nu_3}_{k_1-q,m_3}P^{\nu_4}_{k_2-q,m_4}P^{\nu_5}_{k_1-k_2,m_5},
\end{equation}
$Q(q^2,\alpha)$ and $D(\alpha)$
are, respectively, the third and the second order polynomials in $\alpha$:
\begin{eqnarray}
&&\!\!Q(q^2,\alpha)\!=\overline{Q}(\alpha)q^2\!=[(\alpha_1\!
+\!\alpha_2)(\alpha_3
\!+\!\alpha_4)\alpha_5\!+\!\alpha_1\alpha_2(\alpha_3\!+\!\alpha_4)
\!+\!\alpha_3\alpha_4(\alpha_1\!+\!\alpha_2)]q^2, \\
&&\!\!D(\alpha)=\alpha_5(\alpha_1+\alpha_2+\alpha_3+\alpha_4)+
 (\alpha_1+\alpha_3)(\alpha_2+\alpha_4).
\end{eqnarray}

If, at $q^2=0$,   $J_{\nu_1 \nu_2 \nu_3 \nu_4 \nu_5}(q^2)$ does not 
have singularities then it can be expanded into the Taylor series:
\begin{equation}
J_{\nu_1 \nu_2 \nu_3 \nu_4 \nu_5}(q^2)=\sum_{l=0}^{\infty} J_l~ (q^2)^l.
\end{equation}
The differential operator generating $J_l$ will be a four-fold sum:
\begin{equation}
\overline{Q}^l(\partial)=\sum_{j=0}^{l} \sum_{r=0}^{j} \sum_{i=0}^{j}
\sum_{s=0}^{l-i} \frac{l!~j!~(l-i)!~ \partial_1^{i+s}
\partial_2^{l+s} \partial_3^{l-j} \partial_4^{r+l-j}
\partial_5^{j-i}}
{(l-j)!~(j-i)!~(j-r)!~i!~r!~s!~(l-i-s)!}.
\end{equation}
Applying this operator to the diagram taken at $q=0$ we find:
\begin{eqnarray}
&&J_l=
\sum_{j=0}^{l} \sum_{r=0}^{j} \sum_{i=0}^{j}
\sum_{s=0}^{l-i} \frac{j!~(l-i)!~ 
(\nu_1)_{i+s} (\nu_2)_{l+s} (\nu_3)_{l-j}
(\nu_4)_{r+l-j} (\nu_5)_{j-i}}
{(l-j)!~(j-i)!~(j-r)!~i!~r!~s!~(l-i-s)!}
\nonumber \\
&&~~~~~~\times \int \!\! \int \frac{d^{d+2l}k_1 ~d^{d+2l}k_2}
{[i\pi^{d/2+l}]^2}
~P^{\nu_1+i+s}_{k_1,m_1}P^{\nu_2+l+s}_{k_2,m_2}
P^{\nu_3+l-j}_{k_1,m_3}P^{\nu_4+r+l-j}_{k_2,m_4}
P^{\nu_5+j-i}_{k_1-k_2,m_5}.
\label{Jlgeneric}
\end{eqnarray}
If $\nu_i$  ($i=1,\ldots, 4$ ) are integer,  performing partial
fraction decomposition like
\begin{eqnarray}
&&\frac{1}{(k_1^2-m_1^2+i\epsilon)^{\nu_1}(k_1^2-m_3^2
+i\epsilon)^{\nu_3}}\nonumber \\
&&~~~~~~~~~~~~~~~~=\sum_{j=0}^{\nu_1-1}\frac{(-1)^j~(\nu_3-1+j)!}
{j!~(\nu_3-1)!} \frac{1}{(m_1^2-m_3^2)^{\nu_3+j}
(k_1^2-m_1^2+i\epsilon)^{\nu_1-j}} \nonumber \\
&&~~~~~~~~~~~~~~~~+\sum_{j=0}^{\nu_3-1}\frac{(-1)^{\nu_1}~
(\nu_1-1+j)!}
{j!~(\nu_1-1)!} \frac{1}{(m_1^2-m_3^2)^{\nu_1+j}
(k_1^2-m_3^2+i\epsilon)^{\nu_3-j}},
\label{decomp}
\end{eqnarray}
we can express $J_l$ as a multiple sum each term of which is 
proportional to a bubble integral with only three denominators
raised to some powers. These integrals can be expressed in terms of 
four Appell's function $F_4$ \cite{DT}, which in turn are double 
series. Representation ($\ref{Jlgeneric}$) may be helpful for 
numerical calculations  using the multiple precision program 
\cite{Bailey}.

It took 5 min  on a PC Pentium 90, using FORM \cite{FORM},
to generate in terms of bubble integrals the 30-th coefficient of the
small momentum expansion for the two-loop master diagram with all 
masses  different. Again, as we observed  in the one-loop case,
the number of generated terms in our approach is smaller than in
the direct small momentum expansion. For instance, the 
1-st,2-nd,3-rd,4-th and 5-th  coefficients of the Taylor 
expansion of the two-loop diagram with all masses equal have 
3,6,10,15  and 21 terms,  respectively, in our approach  and 
correspondingly 5,18,45,97 and 182 terms in the other methods 
\cite{OVT2}, \cite{DT}. To reduce the bubble integrals
\begin{equation}
J_{\nu_1\nu_2\nu_3}^{(d+2M)}=\int \!\! \int \frac{d^{d+2M} k_1 
d^{d+2M} k_2}
{[i \pi^{d/2+M}]^2}P^{\nu_1}_{k_1,m_1}P^{\nu_2}_{k_1-k_2,m_2}
P^{\nu_3}_{k_2,m_3},
\end{equation}
encountered in the evaluation of the $M$-th coefficient of the 
expansion, to  the $d=4-2\varepsilon$  dimensional  ones, it was enough
to  use  only one  recurrence relation which is given in \cite{OVT1}:
\begin{equation}
\nu_2\nu_3 d J_{\nu_1~\nu_2+1~\nu_3+1}^{(d+2)}+2\nu_1 m_1^2
J_{\nu_1+1~\nu_2\nu_3}^{(d)}
 -(d-2\nu_1)  J_{\nu_1\nu_2\nu_3}^{(d)} =0.
\end{equation}
Needless to say that in our method the coefficients were expressed
in terms of integrals with the sum of exponents of scalar propagators
larger than in the other methods, i.e.  more complicated integrals. 
Despite of this fact, the sum of those integrals were 
calculated faster. For example, the 6-th coefficient of the 
aforementioned diagram was calculated  two times faster. For  
higher coefficients, our method becomes even more effective comparing
with methods known up to now. Significant improvement in the 
efficiency of the method  may be achieved by implementing
 it as a numerical procedure proposed in \cite{JF}
 using the multiple precision program \cite{Bailey}.

In some simple cases the method proposed  may be helpful in finding
analytic results.  For example, the two-loop
diagram with three denominators:
\begin{equation}
T^{(d)}(q^2,m_1^2,m_2^2,m_3^2)=
\int \!\! \int d^dk_1 d^dk_2~P^{\nu_1}_{k_1,m_1}P^{\nu_2}_{k_1-k_2,m_2}
P^{\nu_3}_{k_2-q,m_3},
\end{equation}
has a rather simple  $Q$ function:
$Q(q^2,\alpha)=\alpha_1 \alpha_2 \alpha_3q^2$.  The coefficients of
the small momentum expansion of the diagram:
\begin{equation}
T^{(d)}(q^2,m_1^2,m_2^2,m_3^2)=\sum_{l=0}^{\infty} \frac{T_l}{l!} 
(q^2)^l
\label{rjad}
\end{equation}
can be written as:
\begin{equation}
T_l=\frac{(-1)^l}{\pi^{2l}}(\nu_1)_l (\nu_2)_l (\nu_3)_l\int\!\!
\int d^{d+2l}k_1
d^{d+2l}k_2~P^{\nu_1+l}_{k_1,m_1}P^{\nu_2+l}_{k_1-k2,m_2}
P^{\nu_3+l}_{k_2,m_3}.
\end{equation}
Using formula (4.3) from Ref. \cite{DT}, we obtain:
\begin{eqnarray}
&&T_l=\frac{\pi^d i^{2-2d}(-1)^l(-m_3^2)^{d-l-\nu_1-\nu_2-\nu_3}}
{\Gamma(\nu_1)\Gamma(\nu_2)\Gamma(\nu_3)\Gamma(\frac{d}{2}+l)}
\nonumber \\
&&~~~~\times \left\{
\Gamma(\frac{d}{2}-\nu_1)\Gamma(\frac{d}{2}-\nu_2)
\Gamma(\nu_1+\nu_2+l-\frac{d}{2})\Gamma(\nu_1+\nu_2+\nu_3+l-d)
\right.
\nonumber \\
&&~~~~~~~~~~~~~~\times F_4(\nu_1+\nu_2+\nu_3+l-d,\nu_1+\nu_2
+l-\frac{d}{2};
\nu_1+1-\frac{d}{2},\nu_2+1-\frac{d}{2}~|~x,y)\nonumber \\
&&~~~~~+y^{\frac{d}{2}-\nu_2}
\Gamma(\frac{d}{2}-\nu_1)\Gamma(\nu_2-\frac{d}{2})
\Gamma(\nu_1+l)\Gamma(\nu_1+\nu_3+l-\frac{d}{2}) \nonumber \\
&&~~~~~~~~~~~~~~\times F_4(\nu_1+l,\nu_1+\nu_3+l
-\frac{d}{2};\nu_1+1-\frac{d}{2},
\frac{d}{2}+1-\nu_2|x,y) \nonumber \\
&&~~~~~ +x^{\frac{d}{2}-\nu_1}\Gamma(\nu_1-\frac{d}{2})
\Gamma(\frac{d}{2}-\nu_2)\Gamma(\nu_2+l)
\Gamma(\nu_2+\nu_3+l-\frac{d}{2}) \nonumber \\
&& ~~~~~~~~~~~~~~\times F_4(\nu_2+l,\nu_2+\nu_3+l-\frac{d}{2};
\frac{d}{2}+1-\nu_1, \nu_2+1-\frac{d}{2}~|~x,y) \nonumber \\
&&~~~~~+x^{\frac{d}{2}-\nu_1}y^{\frac{d}{2}-\nu_2}
\Gamma(\nu_1-\frac{d}{2})
\Gamma(l-\frac{d}{2})\Gamma(\nu_3+l)\Gamma(\frac{d}{2}+l) \nonumber \\
&&\left.
~~~~~~~~~~~~~~
\times F_4(\nu_3+l,\frac{d}{2}+l;\frac{d}{2}+1-\nu_1,
\frac{d}{2}+1-\nu_2~|~x,y)
\right \},
\label{orexF4}
\end{eqnarray}
where $F_4$ is Appell's hypergeometric function, $x=m_1^2/m_3^2$
and $y=m_2^2/m_3^2$.
By inserting ($\ref{orexF4}$) into ($\ref{rjad}$) and using the
series representation for $F_4$ we reproduced the explicit result
given in \cite{Buza}.

At the two-loop level, there are two topologically different
three-point vertex diagrams shown in  Figs. 2a and 2b,
where each line corresponds to a scalar propagator
with an arbitrary exponent.
\begin{center}
\begin{picture}(360,190)(-35,10)
\Line(10,100)(60,100)
\Line(10,150)(60,150)
\Line(10,100)(10,150)
\Line(60,150)(60,100)
\Line(60,100)(35,70)
\Line(35,70)(10,100)
\ArrowLine(35,50)(35,70)
\ArrowLine(10,150)(10,170)
\ArrowLine(60,170)(60,150)

\Text(15,81)[]{$1$}
\Text(57,81)[]{$2$}
\Text(0,125)[]{$3$}
\Text(67,125)[]{$4$}
\Text(35,160)[]{$5$}
\Text(35,107)[]{$6$}
\Text(10,177)[]{$q_1$}
\Text(60,177)[]{$q_2$}
\Text(28,47)[]{$q_3$}
\Text(40,30)[]{$a)$}
\Vertex(10,100){2}
\Vertex(60,100){2}
\Vertex(10,150){2}
\Vertex(60,150){2}
\Vertex(35,70){2}

\Line(120,100)(145,125)
\Line(120,150)(145,125)
\Line(145,125)(170,150)
\Line(145,125)(170,100)
\Line(120,100)(120,150)
\Line(170,150)(170,100)
\Line(145,70)(120,100)
\Line(170,100)(145,70)


\ArrowLine(145,50)(145,70)
\ArrowLine(120,150)(120,170)
\ArrowLine(170,170)(170,150)

\Vertex(120,100){2}
\Vertex(170,150){2}
\Vertex(145,70){2}
\Vertex(120,150){2}
\Vertex(170,100){2}
\Text(125,85)[]{$1$}
\Text(165,85)[]{$2$}
\Text(110,125)[]{$3$}
\Text(177,125)[]{$4$}
\Text(130,150)[]{$5$}
\Text(161,150)[]{$6$}
\Text(120,177)[]{$q_1$}
\Text(170,177)[]{$q_2$}
\Text(137,48)[]{$q_3$}
\Text(150,30)[]{$b)$}

\Line(235,150)(260,70)
\Line(285,150)(260,70)
\CArc(260,143)(25,20,160)
\CArc(260,157)(25,200,340)


\ArrowLine(260,50)(260,70)
\ArrowLine(235,150)(225,170)
\ArrowLine(295,170)(285,150)

\Text(260,178)[]{$3$}
\Text(260,142)[]{$4$}
\Text(235,122)[]{$1$}
\Text(285,122)[]{$2$}

\Vertex(235,150){2}
\Vertex(260,70){2}
\Vertex(285,150){2}

\Text(252,48)[]{$q_3$}

\Text(225,177)[]{$q_2+q_3$}
\Text(295,177)[]{$q_2$}
\Text(265,30)[]{$c)$}

\Text(125,5)[]{${\rm Fig.2~~Two-loop ~ vertex ~diagrams }$}
\end{picture}
\end{center}
\vspace{1cm}
The $Q(\{s \},\alpha)$ and $D(\alpha)$ polynomials for the  planar
diagram Fig.2a are:
\begin{eqnarray}
\label{Qpl}
&&Q_{pl}(\{s\},\alpha)=
       (\alpha_1 \alpha_3
       + \alpha_1\alpha_6
       + \alpha_2\alpha_3
   + \alpha_3\alpha_6 )\alpha_5 q_1^2 \nonumber \\
&&~~~~~~~~~~~~~~~~+(\alpha_1\alpha_4
       + \alpha_2\alpha_4
       + \alpha_2\alpha_6+ \alpha_4\alpha_6)\alpha_5   q_2^2
       \nonumber \\
&&~~+[\alpha_1 \alpha_2 \alpha_5
+\alpha_6(\alpha_2+\alpha_4)(\alpha_1+\alpha_3)
       + \alpha_1\alpha_3(\alpha_2+\alpha_4)
+ \alpha_2\alpha_4(\alpha_1+ \alpha_3)]~ q_3^2, \\
&&\nonumber \\
&&D_{pl}(\alpha)=(\alpha_1+\alpha_2)(\alpha_3+\alpha_4+\alpha_5)
+\alpha_6(\alpha_1+\alpha_2+\alpha_3+\alpha_4+\alpha_5).
\end{eqnarray}
For the nonplanar  diagram  Fig.2c $Q(\alpha)$ and $D(\alpha)$ are:
\begin{eqnarray}
\label{Qnpl}
Q_{npl}(\{s\},\alpha)\!&=&\!\!
 [\alpha_1 \alpha_5 \alpha_6+ \alpha_2\alpha_3~\alpha_4
+\alpha_3\alpha_5(\alpha_1 + \alpha_2+ \alpha_4+ \alpha_6)]~q_1^2
\nonumber \\
       \!&+&\!\! [ \alpha_1 \alpha_3 \alpha_4
       + \alpha_2 \alpha_5\alpha_6
    +\alpha_4 \alpha_6(\alpha_1 + \alpha_2+ \alpha_3+\alpha_5)] ~ q_2^2
\nonumber \\
 \!&+&\!\! [\alpha_2 \alpha_3 \alpha_6
   + \alpha_1 \alpha_4 \alpha_5
   + \alpha_1 \alpha_2(\alpha_3+\alpha_4+\alpha_5+ \alpha_6)] ~q_3^2,\\
&&\nonumber \\
D_{npl}(\alpha)\!\!&=&\!\!(\alpha_1+\alpha_2)(\alpha_3+\alpha_4
+\alpha_5+\alpha_6)+(\alpha_3+\alpha_5)(\alpha_4+\alpha_6).
\end{eqnarray}

Any two-loop scalar diagram  at zero momenta
$G^{(d)}(\{ 0 \},\{m_k^2\})$ can be written, using partial fraction
decomposition ($\ref{decomp}$),  as a multifold sum over bubble 
integrals with three denominators only. As we already mentioned 
those integrals for 
arbitrary space-time dimension and arbitrary exponents of propagators
are combinations of four Appell functions $F_4$.
Therefore, using ($\ref{EXPANSION}$),  the coefficients of
the Taylor expansion of any two-loop integral
can be written in a closed form as a multifold sum over functions
$F_4$ times product of Euler's $\Gamma$ functions.
Multiple sums will be produced by  $Q^l_j(\partial)$
and partial fraction decomposition of the propagators
with the same momenta and different masses.

The representation of the coefficients in terms of $F_4$ functions
may  not be  of  great use in general.
For the cases when some masses are equal or vanish one can get
simpler representations. If, for example, two masses in the
vacuum integral with three denominators are equal, then the
$F_4$ functions are reduced to the somewhat simpler $_4F_3$ functions
\cite{DT}. If one mass is zero then $F_4$ reduce to $_2F_1$
functions.

As an example, we consider the small momentum expansion w.r.t. to the
subset of scalar invariants $q_3^2$, $q_2q_3$  for the diagram Fig.2c:
\begin{eqnarray}
J(q_2,q_3)&=&\int \!\!\! \int \frac{d^dk_1 d^dk_2}
{[i \pi^{d/2}]^2}~P_{k_1-q_2,m_1}
P_{k_1-k_2,m_2}P_{k_2-q_3,m_3}P_{k_2,m_4} \nonumber \\
&=&\sum_{l=0}^{\infty}\sum_{r=0}^{\infty}
\frac{(2q_2q_3)^l(q_3^2)^r}{l!~ r!}  ~C_{lr}(q_2^2).
\end{eqnarray}
By using ($\ref{DQshrink}$), the polynomial $Q(\alpha)$ for this 
diagram
\begin{equation}
Q(\{q_iq_j \},\alpha)=2(q_2q_3) \alpha_1\alpha_2\alpha_3
+\alpha_1\alpha_2(\alpha_3+\alpha_4)q_2^2
+q_3^2\alpha_3[\alpha_1\alpha_2+\alpha_4(\alpha_1+\alpha_2)],
\end{equation}
can be obtained from ($\ref{Qpl}$) or ($\ref{Qnpl}$)
by setting $\alpha_3=\alpha_4=0$ and then replacing
$\alpha_5 \rightarrow \alpha_3$, $\alpha_6 \rightarrow \alpha_4$.
The coefficients $C_{lr}(q_2^2)$ can be represented as:
\begin{eqnarray}
&&C_{lr}(q_2^2)=(-1)^{l+r}r! ~\sum_{j=0}^r \sum_{s=0}^j
\frac{\partial^{l+r-s}_1
\partial_2^{l+r-j+s}\partial_3^r\partial_4^j}{(r-j)!(j-s)!s!} \\
&&~~~~~~~~~~~~~~~~
\times \int \!\!\! \int \frac{ d^{d+2r+2l}k_1 d^{d+2r+2l}k_2}
{[i \pi^{d/2+l+r}]^2}
~P_{k_1-q_2,m_1}P_{k_1-k_2,m_2}P_{k_2,m_3}P_{k_2,m_4}.
\nonumber
\end{eqnarray}
By differentiating and then performing partial fraction decomposition,
the coefficients $C_{lr}(q_2^2)$ will be
expressed as  combinations of integrals:
\begin{equation}
J^{(d+2r+2l)}_{\nu_1 \nu_2 \nu_3}=\int \!\!\! \int \frac{
d^{d+2r+2l}k_1~d^{d+2r+2l}k_2}{ [i \pi^{d/2+l+r}]^2 }
P_{k_1-q_2,m_1}^{\nu_1}P_{k_1-k_2,m_2}^{\nu_2}P_{k_2,m_j}^{\nu_3},
\label{orexint}
\end{equation}
where $m_j$ corresponds to $m_3$ or $m_4$.
To reduce $d+2r+2l$ dimensional integrals to $d$ dimensional 
the generalized recurrence relations proposed in \cite{OVT1}
must be used. Just for illustration we derive several relations
of this kind. In particular one may get the relation:
\begin{eqnarray}
&&\nu_1 q_{2 \mu} J^{(d)}_{\mu,~\nu_1+1 ~\nu_2\nu_3}+(d-\nu_1-\nu_2-\nu_3)
J^{(d)}_{\nu_1 \nu_2 \nu_3}
 \nonumber \\
&&~~~~~~~~~~~~~~~
-m_1^2 \nu_1 J^{(d)}_{\nu_1+1~\nu_2 \nu_3}
-m_2^2 \nu_2 J^{(d)}_{\nu_1 \nu_2+1~\nu_3}
-m_3^2 \nu_3 J^{(d)}_{\nu_1 \nu_2 \nu_3+1}=0,
\label{orex}
\end{eqnarray}
where
\begin{equation}
J^{(d)}_{\mu,~\nu_1 \nu_2 \nu_3}=
\int \!\!\! \int
d^{d}k_1~d^{d}k_2~P_{k_1,m_1}^{\nu_1}P_{k_1-k_2,m_2}^{\nu_2}
P_{k_2,m_3}^{\nu_3}~k_{1\mu}.
\end{equation}
Following the procedure described in \cite{OVT1} the integral
$J^{(d)}_{\mu,~\nu_1\nu_2\nu_3}$ can be expressed in terms of 
a scalar integral
with the space-time dimension $d+2$:
\begin{equation}
J^{(d)}_{\mu,\nu_1+1~ \nu_2 \nu_3}=q_{2 \mu} \partial_2
\partial_3 J^{(d+2)}_{\nu_1+1~ \nu_2 \nu_3}.
\label{vect}
\end{equation}
By inserting ($\ref{vect}$) into ($\ref{orex}$) we obtain:
\begin{eqnarray}
&&\nu_1 \nu_2 \nu_3 q_2^2 J^{(d+2)}_{\nu_1+1~ \nu_2+1~ \nu_3+1}
=(\nu_1+\nu_2+\nu_3-d) J^{(d)}_{\nu_1 \nu_2 \nu_3}
\nonumber \\
&&
~~~~~~~~~~~~~~~~~~~~~~~~
+m_1^2 \nu_1 J^{(d)}_{\nu_1+1~ \nu_2 \nu_3}
+m_2^2 \nu_2 J^{(d)}_{\nu_1 \nu_2+1~ \nu_3}
+m_3^2 \nu_3 J^{(d)}_{\nu_1 \nu_2 ~\nu_3+1}
.
\label{eq49}
\end{eqnarray}
From the equation connecting $d$ and $d-2$ dimensional integrals
given in  \cite{OVT1}, which in our case reads 
\begin{equation}
J^{(d-2)}_{\nu_1 \nu_2 \nu_3}=(\partial_{1}\partial_{2}
+\partial_{1}\partial_{3}+\partial_{2}\partial_{3})
J^{(d)}_{\nu_1 \nu_2 \nu_3}
\end{equation}
another relation follows:
\begin{equation}
J^{(d-2)}_{\nu_1 \nu_2 \nu_3}-\nu_1\nu_2J^{(d)}_{\nu_1+1~ \nu_2+1~
\nu_3}
-\nu_1 \nu_3 J^{(d)}_{\nu_1+1~ \nu_2 \nu_3+1}-
\nu_1 \nu_3J^{(d)}_{\nu_1 \nu_2+1~ \nu_3+1}=0.
\label{eq52}
\end{equation}
By using  Eqs. ($\ref{eq49}$) and ($\ref{eq52}$), several others
obtained in a similar manner  and relations
derived from the method of integration by parts \cite{CT}, one can
reduce the integrals ($\ref{orexint}$) to the set of
$d=4-2\varepsilon$ dimensional master integrals.

As a final example we  consider the  small momentum expansion of the
two-loop scalar vertex diagram \cite{FT} occurring in the
process $H \rightarrow 2 \gamma$:
\begin{equation}
I=\int \!\!\! \int \frac{ d^dk_1~ d^dk_2}{[i \pi^{d/2}]^2}~
P_{k_1+q_1,m}P_{k_1+q_2,m}
P_{k_2+q_1,m}P_{k_2+q_2,m}P_{k_1-k_2,0}P_{k_2,m},
\end{equation}
with $q_1^2=q_2^2=0$. For this kinematics the method of integration
by parts gives:
\begin{equation}
\label{Hggrec}
I=\frac{2}{(d-4)}
\int \!\! \int \frac{d^dk_1~ d^dk_2}{[i \pi^{d/2}]^2}
P_{k_1+q_1,m}P_{k_1+q_2,m}^2 P_{k_2+q_1,m}P_{k_2,m}
\left[P_{k_2+q_2,m}
-P_{k_1-k_2,0}\right].
\end{equation}
The first term is just the product of factorized one-loop integrals
which are known in terms of hypergeometric functions
\cite{BoDa}. By using our method we expand the second integral 
in ($\ref{Hggrec}$)  w.r.t. $q_3^2=(q_1-q_2)^2$ . The $Q$ function
can be obtain from ($\ref{Qpl}$) by setting $q_1^2=q_2^2=0$ and
$\alpha_4=0$:
\begin{equation}
Q(q_3^2,\alpha)=Q_3(\alpha)q_3^2=\alpha_2[\alpha_1(\alpha_3+\alpha_5)
+\alpha_6(\alpha_1+\alpha_3)]q_3^2.
\end{equation}
The differential operator for the $l$-th term of the expansion will 
be:
\begin{equation}
Q_3^l(\partial)=\frac{l!}{\pi^{2l}}~\partial_2^l~
\sum_{j=0}^l \sum_{i=0}^{l-j} \sum_{k=0}^j
\frac{\partial_1^{l-j+k}\partial_3^{l-k-i}\partial_5^i\partial_6^j}
{(l-j-i)!~(j-k)! ~i!~k!}.
\end{equation}
By applying this operator to the integral
$$
\int \!\! \int \frac{d^{d+2l}k_1~d^{d+2l}k_2}{[i \pi^{d/2}]^2}
P_{k_1,m_1}P_{k_2,m_2}
P_{k_1,m_3}P_{k_1,m_5}P_{k_1-k_2,m_6},
$$
setting $m_1=m_2=m_3=m$,  $m_6=0$ and adding the contribution from
the factorized integrals we obtain the series expansion of $I$ in
$q_3^2=(q_1-q_2)^2$:
\begin{equation}
I=m^{12-2d} \sum_{l=0}^{\infty} I_l \left( \frac{q_3^2}{m^2} \right)^l,
\end{equation}
where
\begin{eqnarray}
&&I_l=\frac{2(l+1)!}{(d-4)} \left[
\frac{\Gamma^2(\frac12)}{(l+1)!~4^{l+2}} \sum_{j=0}^l
\frac{\Gamma(l-j+3-\frac{d}{2})\Gamma(j+3-\frac{d}{2})}
{\Gamma(l-j+\frac32)\Gamma(j+\frac32)~(j+1)}
+\frac{\Gamma(l+6-d)}{\Gamma(\frac{d}{2}+l)}\right.
 \nonumber \\
&&\\
&&~~~~\times \left.
\sum_{j=0}^l \frac{j!~\Gamma(\frac{d}{2}+l-j-1)}
{ \Gamma(l+j+7-d)}
 \sum_{k=0}^j \frac{(l-k)!~\Gamma(k+3-\frac{d}{2})
\Gamma(l+j-k+4-\frac{d}{2})}
{(l-j)!~(j-k)!~(k+1)!~(2l-k+2)!}\right].
\nonumber
\end{eqnarray}
For $d=4$ this formula confirms the coefficients given in \cite{FT}.

\section{Conclusions}
We presented a new systematic method for the momentum expansion
of  arbitrary scalar multiloop Feynman diagrams. The method can be
used for the small momentum expansion and  as an ingredient for
the large momentum expansion or other kind of asymptotic
expansions. An important feature of our
method is that it can be used in cases when only some momenta
are small and the  others arbitrary.
This extends the applicability of our method  to a wider class  of
physical problems.
Of special interest will be the application  of our method to the
evaluation of the moments of structure functions in deep inelastic
scattering.

\vspace{0.5cm}
{\Large
{\bf Acknowledgments }
}
\vspace{0.2cm}

I am  grateful to J. ~Fleischer, F.~Jegerlehner and R.~Pittau for 
useful discussions and for carefully reading the manuscript.

\end{document}